\documentclass[journal]{IEEEtran}
\usepackage{xcolor}

\usepackage{subcaption}

\usepackage{algorithmic}
\usepackage{graphicx}
\usepackage{amsmath}
\usepackage{textcomp}
\usepackage{xcolor}
\def\BibTeX{{\rm B\kern-.05em{\sc i\kern-.025em b}\kern-.08em
    T\kern-.1667em\lower.7ex\hbox{E}\kern-.125emX}}

\usepackage{adjustbox}
\usepackage{multirow}
\usepackage{amssymb,mathtools}

\usepackage{array}
\newcolumntype{P}[1]{>{\centering\arraybackslash}p{#1}}
\newcolumntype{M}[1]{>{\centering\arraybackslash}m{#1}}

\usepackage{tikz}

\usepackage{xcolor}
\usepackage[linesnumbered,ruled,vlined]{algorithm2e}

\SetCommentSty{mycommfont}

\SetKwInput{KwInput}{Input}                
\SetKwInput{KwOutput}{Output}              

\usepackage{graphicx}

\ifCLASSINFOpdf
\else
\fi
\begin{document}
%
\title{Adaptive Hybrid Heterogeneous IDS for 6LoWPAN}
%
%
%


        
\author{	
 
Aryan Mohammadi Pasikhani,
        John A Clark,
        Prosanta Gope, \IEEEmembership{Senior Member, IEEE}

\IEEEcompsocitemizethanks{\IEEEcompsocthanksitem Aryan Mohammadi Pasikhani, John A Clark, Prosanta Gope are with Department of Computer Science, University of Sheffield, Regent Court, Sheffield S1 4DP, United Kingdom.
(E-mail: amohammadipasikhani1@sheffield.ac.uk)

\textbf{Corresponding author:} Dr. Aryan Mohammadi Pasikhani
}
}

\maketitle

\begin{abstract}
IPv6 over Low-powered Wireless Personal Area Networks (6LoWPAN) have grown in importance in recent years, with the Routing Protocol for Low Power and Lossy Networks (RPL) emerging as a major enabler. However, RPL can be subject to attack, with severe consequences. Most proposed IDSs have been limited to specific RPL attacks and typically assume a stationary environment. In this article, we propose the \emph{first} adaptive hybrid IDS to efficiently detect and identify a wide range of RPL attacks (including DIO Suppression, Increase Rank, and Worst Parent attacks, which have been overlooked in the literature) in evolving data environments. We apply our framework to networks under various levels of node mobility and maliciousness. We experiment with several incremental machine learning (ML) approaches and various ‘concept-drift detection’ mechanisms (e.g. ADWIN, DDM, and EDDM) to determine the best underlying settings for the proposed scheme. 
\end{abstract}


%
\IEEEpeerreviewmaketitle

\begin{table*}[t!]
\caption{Related Works}

\scalebox{.83}{
\begin{tabular}{|c|p{0.4\linewidth}|p{0.4\linewidth}|c|c|c|c|c|}
\hline
\textbf{Scheme}&\textbf{Method}&\textbf{Attacks Considered}&\multicolumn{5}{|c|}{\textbf{Desirable Properties}} \\
\cline{4-8} 
\textbf{} & \textbf{} & \textbf{} & \textbf{\textit{DP1}}& \textbf{\textit{DP2}}& \textbf{\textit{DP3}} & \textbf{\textit{DP4}}
& \textbf{\textit{DP5}} 
\\
\hline

\cite{raza2013svelte} & Active decentralised IDS
 & SH and GH (using Cooja simulator)
 & $\times$  & $\times$ & $\times$ & $\times$ & \checkmark \\
\hline

\cite{kaliyar2020lidl} & Specification-based IDS & WH and Sybil (using Cooja simulator)
 & $\times$  & $\times$ & $\times$ & $\times$ & $\times$ \\
\hline

\cite{shafique2018detection} & Specification-based active centralised IDS & SH (using Cooja simulator)
 & $\times$  & $\times$ & $\times$ & $\times$ & $\times$ \\
\hline

\cite{9415869} & Hybrid specification-based IDS & SH, BH, GH, DA, WH, Clone Id, Replay, Sybil, Rank, VN and WP (Netsim)
& $\times$ & $\times$ & $\times$ & \checkmark & $\times$  \\
\hline




\cite{shukla2017ml} & Hybrid ML-based IDS & WH
& $\times$  & \checkmark & $\times$ & \checkmark & $\times$ \\
\hline

\cite{foley2020employing} & Ensemble Voting (MLP and RF)  & SA, VN, SH, and BH 
& $\times$ & $\times$ & $\times$ & $\times$ & $\times$  \\
\hline


\cite{bostani2017hybrid} & Unsupervised Optimum-Path Forest Clustering & SH, WH, and SF
& $\times$ & $\times$ & $\times$ & $\times$ & $\times$ \\
\hline

\cite{napiah2018compression} & Hybrid ML-IDS using passive monitoring technique & SH, WH, and DA (using Cooja simulator)
& $\times$  & $\times$ & $\times$ & $\times$ & $\times$ \\
\hline

\cite{shreenivas2017intrusion} & Active decentralised hybrid IDS  & SH (using Cooja simulator)
 & $\times$ & $\times$ & $\times$ & $\times$ & $\times$ \\
\hline

\cite{farzaneh2019anomaly} & Active decentralised anomaly-based IDS &  DA and NA
 & $\times$ & $\times$ & $\times$ & $\times$ & \checkmark \\
\hline

\cite{kasinathan2013ids} & Passive decentralised signature-based IDS & DA (using Cooja simulator)
 & $\times$ & $\times$ & $\times$ & $\times$ & $\times$ \\
\hline

\cite{le2016specification} & Active decentralised specification-based & WP, DA, SH, and DF
& $\times$ & $\times$ & $\times$ & \checkmark & $\times$\\
\hline

\cite{kareem2021ml} & Online adaptive RF + concept drift & KDDCup99 (application layer attacks)
 & $\times$ & $\times$ & $\times$ & $\times$ & $\times$ \\
\hline

\cite{martindale2020ensemble} & Online RF (Hoeffding Trees) & KDDCup99 (application layer attacks)
& $\times$ & $\times$ & $\times$ & $\times$ & $\times$\\
\hline

\cite{li2018ai}& Ensemble Weighted Voting, RF & KDDCup99 (application layer attacks)
& $\times$ & D/N & D/N & D/N & D/N \\
\hline

\cite{yuan2018concept} & Concept drift (HDDM) based ensemble incremental learning approach in IDS & KDDCup99 (application layer attacks)
& \checkmark & D/N & D/N & D/N & D/N \\
\hline

\cite{singh2015intrusion} & Online Sequential-Extreme Learning Machine (OS-ELM) &  NSL-KDD 2009 (application layer attacks)
& \checkmark & D/N & D/N & D/N & D/N \\
\hline

Our Scheme & One-Class SVM, incremental OzaBaggingADWIN using KNN, and HalfSpace-Trees & SH, BH, GH, DA, DS, IR, WH, and WP (Netsim v13)
& \checkmark& \checkmark & \checkmark &\checkmark & \checkmark \\
\hline


\multicolumn{8}{l}{$^{\mathrm{*}}$D/N: Different Network-technology. $^{\mathrm{*}}$
In the “Attack” column, the later entries refer to available datasets that contain a variety of attacks, (but these exclude RPL attacks); 
}
\\

\multicolumn{8}{l}{\text{\checkmark: Satisfy; $\times$: Not addressed; \checkmark$^{\mathrm{*}}$: Satisfy part of that desirable property; \textbf{SH:} Sinkhole, \textbf{BH:} Blackhole; \textbf{GH:} Grayhole;  \textbf{DA:} DIS Flooding; \textbf{IR:} Increase Rank; \textbf{WH:} Wormhole; }
}\\
\multicolumn{8}{l}{\text{\textbf{DS:} DIO Suppression; \textbf{WP:} Worst Parent; \textbf{DP1}: Adaptive intrusion detector; \textbf{DP2}: Lightweight; \textbf{DP3}: Accurate in evolving data environment; \textbf{DP4}: Detect a wide range of RPL attacks;}}\\
\multicolumn{8}{l}{\text{\textbf{DP5}: Detect known and unknown (a.k.a unseen) intrusions;}
}\\
\end{tabular}
}

\label{tab:Related Works}%
\end{table*}

\section{Introduction}
\label{sec:Introduction}

Internet of things (IoT) networks are generally Low-Power and Lossy Networks (LLNs) consisting of heterogeneous devices with limited power, memory, and processing resources. 
LLNs have been deployed in various sectors such as agriculture, control, the built environment and rural environment monitoring \cite{9383263}. For efficient routing in LLNs, the Internet Engineering Task Force (IETF) introduced the Routing Protocol for Low-Power and Lossy Networks (RPL) \cite{rfc6550}. Global connectivity, resource constraints and RPL vulnerabilities expose 6LoWPAN to various routing threats, internally (within the 6LoWPAN) and externally (through the Internet). Existing routing attacks (e.g. Blackhole, Grayhole, Wormhole, and DODAG Informational Solicitation (DIS) flooding attacks) \cite{9383263} cause the RPL to generate suboptimal routing topologies, isolate legitimate nodes, and cause significant overheads over the target network and nodes.

To deal with the security threats in RPL, a variety of Intrusion Detection System (IDS) proposals have been introduced in the literature. A network-based IDS can identify threats by analysing sniffed packets. IDSs can be signature-based, anomaly-based, specification-based, or hybrid \cite{9383263}. Signature-based IDSs use known signatures of  attacks  to identify intrusions. They can classify known intrusions accurately but  require huge storage space to maintain the database of reference signatures which must be updated continually. They cannot reliably detect hitherto unseen intrusions. Anomaly detectors build a profile of normal behaviour and detect significant deviations from that normal profile.
Although anomaly-based IDS requires less storage space to identify abnormal instances \cite{raza2013svelte}, it is prone to generate many false-positive (FP) classifications (i.e., identifying legitimate activity as anomalous). Specifcation based approaches typically detect deviational behaviour from a formalised specification, e.g. that provided by a protocol description. 
The hybrid detection strategy combines  existing detection strategies to incorporate their strengths and minimise their downsides. Various approaches to intrusion detection are found in the literature, e.g. statistical, rule-based and machine learning-based. There are three major categories of ML-based IDS (ML-IDS): supervised (having access to labelled  normal and malicious data instances), unsupervised (without access to any labelled data), or semi-supervised (where not all data is labelled, or else access is restricted to normal instances \cite{bhuyan2012survey}). 

The 6LoWPAN has a streaming data environment. An IDS does not have access to the entire data stream  and cannot afford to store all incoming data instances. Existing IDSs proposed for 6LoWPAN  work only in  stationary environments where the number of nodes in each scenario does not change. However, 6LoWPAN has an evolving data environment, where node movement, inaccessibility, changes in running applications, and unforeseen attacks alter the data stream distribution. 6LoWPAN nodes cannot store a large volume of data. Moreover, in non-stationary evolving environments, the data distribution evolves unpredictably and so the system needs to update its model incrementally or retrain it using recently observed batches of data. To address the aforesaid issues, “\textit{concept drift}” detection approaches have been introduced in different network paradigms to enable adaptivity of the IDS \cite{gama2014survey}. A  “\textit{concept}” can be defined as a joint distribution $P(X | Y)$, where $X$ denotes a vector of attribute values (features) and $Y$ is the target value (label) \cite{webb2016characterizing}. Concept drift is a shift in the data distribution $P(X)$, where $P_{t}(X, Y) \neq P_{t+1}(X, Y)$. Thus, over time, the likelihood that observed data indicates normal system operation may change, e.g. if new malware has been crafted, or has otherwise adapted, to ‘look like’ benign software. The rate of concept drift is unknown to the system and can be abrupt, incremental, gradual or recurring \cite{gama2014survey}. Concept-drift Detection (CD) methods can enable an IDS to adapt to unforeseen intrusions and identify shifts in the network data stream \cite{bhuyan2012survey}. Additionally, CD approaches use storage and memory resources efficiently and facilitate fast classification.

Developing an adaptive IDS capable of accurately classifying the 6LoWPAN evolving data stream is a challenging task. The classifier needs to update itself with each change (shift) in the environment to continue to detect novel  attacks. Re-training a classifier using the entire training data is computationally expensive and generally infeasible. This article proposes the use of streaming data mining techniques and drift detection to provide a novel adaptive form of hybrid ensemble capable of enhancing system performance. The proposed scheme can identify various routing attacks. Internal  attacks (sourced inside 6LoWPAN) include sinkhole, blackhole, and grayhole attacks. External (sourced over the Internet) attacks include wormhole and DIS flooding attacks.

Different ensembling techniques have been adopted and compared in this article. A passive decentralised monitoring technique (where anomaly-based IDS agents passively monitor network communications and send abnormal/suspicious observations to the central IDS for further analysis) is used to collect and monitor LLN traffic from different locations and avoid additional computational overheads over legitimate nodes for intrusion detection purposes.

\subsection{Desirable Properties}
\label{subsec:Desirable_Properties}
Our proposed IDS approach aims to achieve the following Desirable Properties (DPs).

\begin{itemize}


\item \textbf{DP1:} the IDS should be able to identify routing attacks  in an evolving data stream environment by updating its detection model when drift is detected.

\item \textbf{DP2:} the IDS should not need excessive memory  and computational resources whilst being  able to identify  routing attacks precisely.


\item \textbf{DP3:} the IDS should work over 6LoWPAN networks incorporating mobile nodes.


\item \textbf{DP4:} the IDS should be able to detect a wide range of RPL attacks. 

\item \textbf{DP5:} the IDS should be able to detect both known and previously unseen intrusions.

\end{itemize}

\subsection{Related Work}
\label{sec:Related_Works}
A broad range of routing vulnerabilities in 6LoWPAN and the lack of effective built-in security mechanisms in RPL \cite{9383263} have encouraged researchers to develop IDSs  for detecting RPL attacks. 
Various monitoring and detection strategies \cite{9383263} have been considered. These  \cite{kaliyar2020lidl,pongle2015real,mayzaud2016using,shafique2018detection} typically use a specification-based IDS to detect Sinkhole (SH), Wormhole (WH) and DIS flooding (DA) attacks.  54\% of  existing IDSs employed a specification-based detection strategy for detecting routing attacks in 6LoWPAN \cite{9383263}. Specification-based IDSs employ a set of static rules for identifying intrusions; they cannot update their rules automatically. Only 21\% of reported works  have considered a hybrid detection strategy  \cite{9383263} but none considers  mobility of nodes.

The shortcomings of the statistical and rule-based detection approaches \cite{9383263} have encouraged researchers to apply machine learning (ML) algorithms to enhance the performance of IDS in 6LoWPAN. Among  existing hybrid IDSs, only a few  \cite{shukla2017ml,foley2020employing,bostani2017hybrid} are ML-based. Moreover,
they \cite{foley2020employing,shukla2017ml,napiah2018compression,bostani2017hybrid} use offline ML approaches, where the intrusion detection model is constructed using a stationary batch of training data. The  batch-trained ML-IDS degrades as the data stream environment evolves \cite{bhuyan2012survey}.
Nevertheless, legitimate 6LoWPAN nodes  often have limited memory and cannot store extensive records of malicious activities. This inevitably means that less critical records should be replaced with vital ones over time. To the best of our knowledge, no existing IDS for 6LoWPAN does this.

Various proposed monitoring techniques observe  inter-node communication in the 6LoWPAN \cite{9383263} (e.g. centralised and decentralised active or passive monitoring approaches). They \cite{kaliyar2020lidl,pongle2015real,shafique2018detection,shreenivas2017intrusion,raza2013svelte,farzaneh2019anomaly,foley2020employing,shukla2017ml,bostani2017hybrid} employ an active monitoring technique to detect RPL attacks. According to \cite{9383263}, $\sim$77\% of existing IDSs used an active monitoring technique, where legitimate nodes were required to participate in intrusion detection tasks with centralised or decentralised intrusion detectors. Active monitoring can provide more information about node configuration (e.g. geographical location, energy consumption, and CPU, RAM, ROM usage) and result in more accurate detection of RPL attacks. However, it also causes additional computational overhead on the legitimate nodes. Consequently, some 6LoWPAN IDS papers employ passive centralised \cite{napiah2018compression,viegas2018reliable} and passive decentralised \cite{kasinathan2013ids,mayzaud2016using,mayzaud2017distributed,mohammadi2021reinforcement}  approaches. Passive monitoring  does not cause any additional computation overhead for legitimate nodes \cite{mayzaud2017distributed}. 
Nevertheless, it  can provide IDS only with control packets that are multicasted or unicasted by monitoring nodes' neighbours.

According to \cite{9383263}, existing IDS mainly focus on detecting sinkhole (21\%), grayhole (14\%), blackhole (10\%) and DIS flooding (10\%) attacks while other RPL attacks are overlooked.
No research in the literature examines the performance of IDS against external routing attacks (external DA and WH), and there is no research detecting DS (DIO Suppression) and IR (Increase Rank) attacks \cite{9383263}.
Furthermore, only 13\% of RPL IDS research has considered mobility \cite{9383263}. Table \ref{tab:Related Works} shows the related works in the literature and the contributions that this article makes.


\subsection{Motivation and Contribution}
\label{sec:Contribution}

The RPL is vulnerable to various routing threats (e.g. Sinkhole, Blackhole, and Wormhole). Further more, the 6LoWPAN data environment evolves on an unpredictable basis.
Different IDSs have been proposed in the literature to detect existing RPL attacks in 6LoWPAN (as discussed in Section \ref{sec:Related_Works}). However, none of the existing IDS satisfies all the desirable properties (as mentioned in Section \ref{subsec:Desirable_Properties}). 
In 6LoWPAN, an IDS observes a considerable (unbounded) volume of data as a continuous flow; hence, it cannot explicitly store all observations to identify anomalous activities.
To maintain detection performance, it is expected that the IDS modify its detection model on a regular basis and incrementally adapt to unforeseen data distributions.
This article proposes and evaluates an adaptive heterogeneous ensemble hybrid IDS framework to detect various types of RPL attacks in 6LoWPAN. The hybrid detection strategy helps the proposed framework to balance the computational cost of the anomaly-based intrusion detection and the storage cost of the signature-based intrusion detection over legitimate nodes. Besides, various incremental ML algorithms and ensemble techniques are evaluated to determine the most suitable combinations for the proposed system. The major contributions of this article are:
\begin{itemize}


\item A new adaptive hybrid IDS to detect internal and external RPL attacks. 

\item An efficient concept-drift-based ML-IDS, maintaining effectiveness in the face of environmental change.

\item An effective approach to identifying a wide range of RPL attacks, including less  researched ones.

\item An IDS which is resilient against known and previously unseen RPL intrusions.

\item A comprehensive and publicly available dataset for ML-based IDSs containing a extensive range of RPL attacks.

\end{itemize}

\subsection{Organisation}

\label{sec:Organisation}
The rest of the article is organised as follows. In Section \ref{sec:Proposed Scheme}, we present our proposed scheme. In Section \ref{sec:Implementation and Evaluation}, we describe our implementation and evaluation details. Section \ref{sec:Conclusion} concludes the paper.

\section{Proposed Scheme}
\label{sec:Proposed Scheme}

\begin{figure*}[t]
\center{\includegraphics[width=180mm]{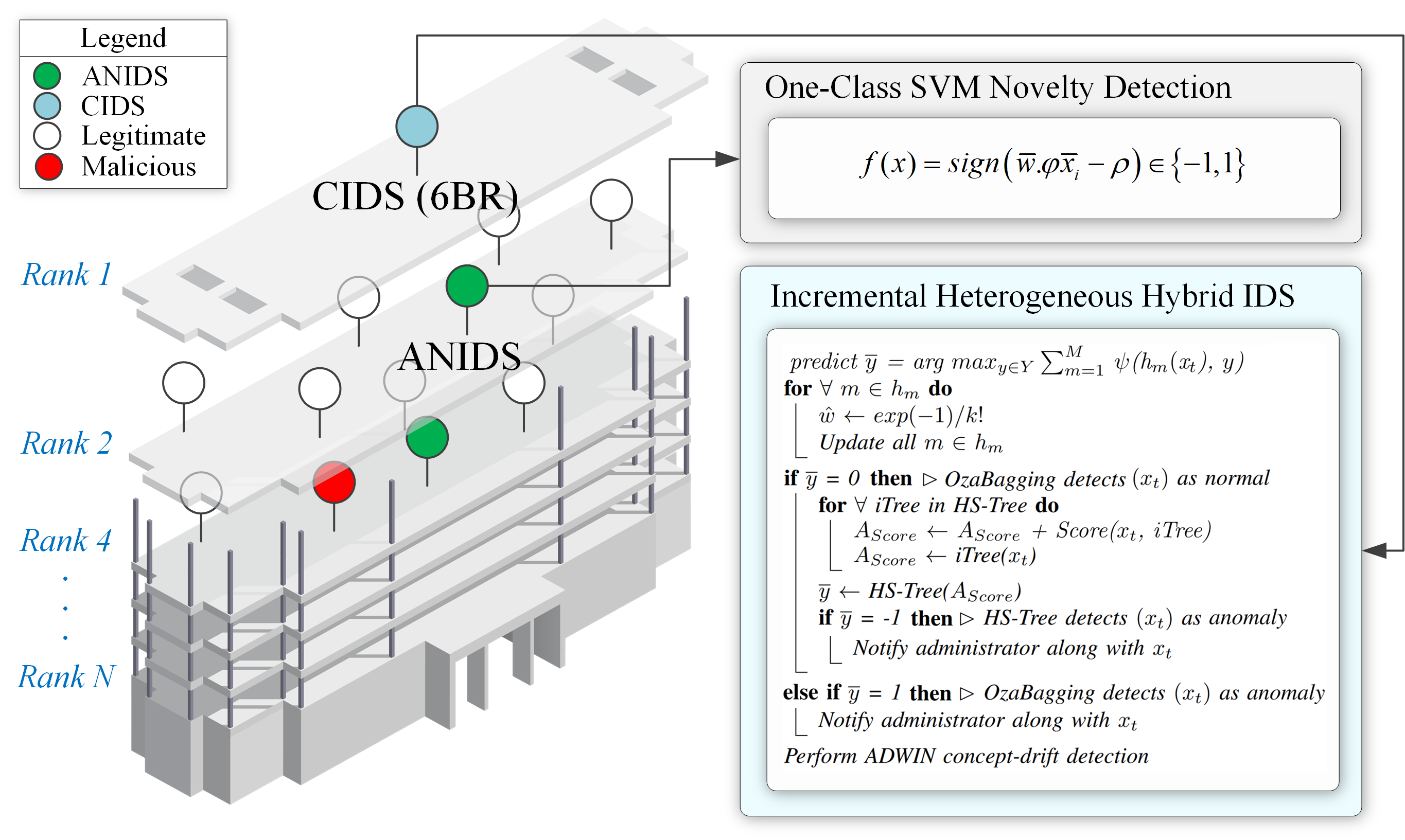}} 
\caption{System Model.}
\label{fig:framework} 

\end{figure*}

Our proposed scheme employs a passive decentralised monitoring approach (readers may refer to \cite{mayzaud2017distributed} for more details) using a cluster-based placement \cite{monitoring_clusterhead} strategy to
analyse the data stream in 6LoWPAN. Anomaly-based detectors are spread over the 6LoWPAN to analyse their neighbours' control packets and report abnormalities to the Centralised IDS (CIDS) on the 6LoWPAN Border Router (6BR). The CIDS is an adaptive heterogeneous hybrid IDS that protects 6LoWPAN against internal and external intrusions. 
Fig. \ref{fig:framework} illustrates the system architecture. The proposed scheme has three components: an anomaly-based network IDS (ANIDS) (Section \ref{sec:ANIDS}), an incremental ensembles of signature-based IDSs (Section \ref{sec:ozabag_lit}), and incremental ensembles of anomaly-based IDSs (Section \ref{sec:hs_tree}) (described below). Algorithm \ref{Alg:proposed_algorithm_ch2} shows the proposed scheme.

\begin{algorithm}[t!]
\DontPrintSemicolon
    \small


  \textbf{\textit{Initialisation}}
  

  \textit{
  A stream of pair $(x, y)$, as $(x_{0},y_{0}), (x_{1},y_{1}) ...$ $(x_{T},y_{T})$, arriving
one-by-one over time.
  }\\

  \textit{X is an evolving data stream (X $\rightarrow$ $\infty$), where $x_{t}$ is  a set of features observed at time $t$ (now).\\ $y$ is the real class label and $\overline{y}$ is the classifier prediction. $Y is\ \{-1, 1\}$}
  
  \textit{$C_{A}$: $C_{OCSVM}$ $\cup$ $C_{HST}$ // Anomaly Classifiers.}
  
  \textit{$C_{OCSVM}$: One-class SVM Classifiers $\subseteq$ $C_{A}$.}
  
  \textit{$iTree$: a HalfSpace-Tree.}
  
  \textit{$\omega$: Window Size.}

  \textit{$A_{Score}$: Anomaly Score.}
   
  \textit{$C_{HST}$: HalfSpace-Trees ensemble classifier $\in$ $C_{A}$.}

\textit{M is the number of models in the ensemble.}

\textit{$h_{m}$ is an adaptive OzaBagging ensemble model induced by learners $m$ $\in$ $\{m_{1} ... m_{n}\}$.}

  
  
   \textit{Count $\leftarrow$ 0.}
   
   \textit{r: mass profile of a node in the reference window. //mass is used as a measure to rank anomalies.}

   \textit{l: mass of a node in the latest window.}

   
   \textit{$k$: Generate poisson $(\lambda=1)$}

   \textit{$\psi$: is the generalised Kronecker function: $\psi(a, b)$ is 1 if a == b, and 0 otherwise.}
    \vspace{1mm}

 \hrule
 


  


  \For(\tcp*[h]{\textit{upon receiving an input ($x_{t}$)}}){each $(x_{t})$ in $X$}  
  {
  \textit{$\delta \leftarrow$ using Eq. \ref{eq:OCSVM_attackdetection} $c$ classifies $(x_{t})$, where $c$ $\in$ $C_{OCSVM}$}\\
  \If(\tcp*[h]{\textit{$c$ has classified $(x_{t})$ as malicious}}){$\delta$ == -1} 
  {
    \textit{ predict $\overline{y}$ = arg $max_{y \in Y}\sum_{m=1}^{M}$ $\psi$($h_{m}$($x_{t}$), $y$)}
    
    \For{all $m$ $\in$ $h_{m}$}
    {
        \textit{$\hat{w}$ $\leftarrow$ $exp(-1)$$/$$k!$ }\\
        \textit{Update $m$ with $(x_{t}, y_{t})$ and weight $\hat{w}$}
    }
    \If(\tcp*[h]{\textit{$h_{m}$ detect $(x_{t})$ as normal}}){$\overline{y}$ == -1}
    {
        \textit{$A_{Score}$ $\leftarrow$ 0}
        
        \For{all iTree in $C_{HST}$}
        {
          \textit{$A_{Score}$ $\leftarrow$ $A_{Score}$ + Score($x_{t}$, iTree) // accumulate scores}
          
          \textit{UpdateMass($x_{t}$, iTree.root, false) // update mass l in iTree }
        }
        
        \textit{Report $A_{Score}$ as the anomaly score for $x_{t}$}
        
        \textit{Count++}

        \If{Count == $\omega$}
        {
        \textit{Update model : Node.r $\leftarrow$ Node.l for every node with non-zero mass r or l}

        \textit{Reset Node.l $\leftarrow$ 0 for every node with non-zero mass l}

        \textit{Count $\leftarrow$ 0}
        }

    }
    \If{ ADWIN detects change in error of one of the models $(h_{m})$}
    {
    \textit{Replace the model with highest error with a new model}
    }

  }
  \textit{\textbf{Output:} Notify administrator if $x_{t}$ is anomalous}
  }
\caption{Proposed Algorithm}
\label{Alg:proposed_algorithm_ch2}

\end{algorithm}
\vspace{-6mm} 

\subsection{Anomaly-based network IDS} 
\label{sec:ANIDS}
Since the CIDS on the 6BR cannot observe network communications of distant nodes (since the 6BR has limited radio range and RPL may operate in storing mode \cite{9383263}), the proposed scheme distributes Anomaly-based Network IDS (ANIDS) agents to passively monitor multicasted and unicasted control packets of their neighbouring nodes without requiring significant storage space. As shown in Experiment 1 (Section \ref{sec:Experiment One}), a  One-Class SVM (OCSVM) can provide excellent performance in detecting intrusions with negligible false-alarms and excellent recall value.
The OCSVM is a novelty detection algorithm that develops a model of safe activities and classifies instances as an outlier (anomalous) if they deviate from its profile. The outcome of OCSVM is bipolar, $y_{t}=-1$ for $x_{t}\in$ outliers and $y_{t}=+1$ for $x_{t}\in$ inliers. In OCSVM, the classifier assumes that the given training dataset $X$ contains only normal (safe) instances, $X$=$\{\overline{x}_{1}, \overline{x}_{2}, ..., \overline{x}_{N}\}$ $\overline{x}_{i}$ $\in$ $Normal$, and considers the origin of a kernel-based transformed representation as an outlier. OCSVM aims to discover a separating boundary (hyperplane) $\overline{w}.\phi(\overline{x}_{i})$ that maximises the distance between normal instances ($\overline{x}$) and the origin $(0, 0)$, $\overline{w}.\phi(\overline{x}_{i})-\rho=0$ (define the hyperplane) where $\overline{w}$ and $\phi(.)$ denote weight and SVM kernel (a function that projects data into a high dimensional space to increase the discriminatory capability of the classifier) respectively; $\rho$ denotes the maximal margin (threshold), Eq. \ref{eq:OCSVM2}, with $N$ instances ${\overline{x}}_{i\in<1,N>}$. According to \cite{maglaras2014real}, the OCSVM can be solved efficiently using the quadratic Eq. \ref{eq:OCSVM1}. The $\nu$ (Nu)  is upper bounded by the fraction of outliers and lower bounded by the fraction of support vectors. The $\nu$ intends to fine-tune the trade-off between over-fitting and generalisation. The conjoint usage of $\nu$ and the slack variable $\xi$ ($\xi \geq 0$) enables the system to handle a dataset that contains a small fraction of outliers. In other words, $\nu$ is the probability of finding an outlier in $X$, where $outliers$ $\subseteq$ $X$. The $\gamma$ (gamma) determines how much influence a single training example has. The larger $\gamma$ is, the closer other examples must be to be affected. Since it is expected that ANIDS generate some degree of false-positive alarms (wrongly classifying safe instances as intrusions),  the instances that are classified as anomalies will be further analysed by the CIDS.


\begin{equation}
\overline{w}.\phi(\overline{x}_{i}) \geq \rho - \xi_{i}\ \forall \overline{x}_{i} \in \textit{X and } \xi_{i} \geq 0, \forall\ i\in\{1, ..., N\}
\label{eq:OCSVM2}
\end{equation}

\begin{equation}
Min_{\overline{w},\overline{\xi},\rho}=\bigg[\frac{1}{2}\|\overline{w}\|^{2}+\left(\frac{1}{\nu\gamma}\sum_{i=1}^{n}\xi_{i}\right)-\rho\bigg]
\label{eq:OCSVM1}
\end{equation}



\begin{equation}
y_{i} = sign(\overline{w}.\phi(\overline{x}_{i}) - \rho)
\label{eq:OCSVM_attackdetection}
\end{equation}

where the $y_{i}$ in Eq. \ref{eq:OCSVM_attackdetection} is an inliner (+1) if $\overline{w}.\phi(\overline{x}_{i}) - \rho  \geq 0$ and an outlier (-1) otherwise.

\subsection{Central IDS} 
\label{sec:CIDS}

The CIDS contains an incremental heterogeneous hybrid IDS and is responsible for analysing internal and external data streams.
It analyses the external network traffic coming to the 6LoWPAN and internal network communications among LLN nodes. Moreover, an observation that is classified as anomalous by any ANIDS will be reported to CIDS for more in-depth analysis. The CIDS analyses the anomalous observations through its incremental ensemble of signature-based IDS and an incremental ensemble of anomaly-based IDS to make more accurate classifications.
Experiments 2 and 3 (in Section \ref{sec:Experiment Two}) show that the incremental ensemble of OZABagging with KNNADWIN learners and HalfSpace-Trese (HS-Trees) \cite{tan2011fast} create a hybrid IDS that provides excellent performance in detecting intrusions. 
The adaptivity through Concept-drift Detection (CD) enables the framework to enhance its intrusion detection performance over time by adapting to unforeseen intrusions and changes in data distributions. The outcomes of Experiments 4, and 5 show that the adaptive sliding window (ADWIN) CD algorithm \cite{gama2014survey} enhances the performance of the proposed scheme while using limited processing and memory at any point in time.


\subsubsection{Incremental ensemble of signature-based IDSs}
\label{sec:ozabag_lit}

Incremental ensemble classifiers provide better detection performance at the cost of more computation and memory usage \cite{gomes2017survey}. An ensemble classifier $f(c_{1}(x_{t}), c_{2}(x_{t}) ... c_{n}(x_{t}))$ is a set of classifiers $(c_{i})$ that make predictions over a given instance of the feature set $(x_{t})$. The OzaBagging classifier \cite{oza2001online} builds an ensemble of classifiers such that $\forall$ $c_{i}$ $\in$ $C$, $c_{i}$ is trained over different bootstrap instances. Since it is challenging to draw samples with replacement in an online streaming environment, the OzaBagging classifier weights the observed instances using a Poisson in bootstrap replica \cite{bifet2009adaptive}. The OzaBaggingADWIN \cite{oza2001online,bifet2009new} is the OzaBagging algorithm with ADWIN concept-drift detection. The OzaBaggingADWIN implements several ADWIN drift detectors to monitor classifier error rates. On the detection of concept drift, OzaBaggingADWIN replaces the worst classifier $c_{i}$ $\in$ $C$ with a new classifier, a procedure described as “replace the loser” \cite{bifet2009adaptive}.

\subsubsection{Incremental ensemble of anomaly-based IDSs}
\label{sec:hs_tree}

Although adopting adaptivity (concept-drift detection) enables a signature-based IDS to learn unforeseen intrusions (discussed in Section \ref{sec:Adaptivity}), a signature-based IDS is prone to some degree of false-negative alarms for unknown intrusions. To enable the proposed framework to identify unknown intrusions, the HalfSpace-Trees (HS-Trees) algorithm \cite{tan2011fast} analyses observations that are classified as normal so far. In HS-Trees, each tree contains nodes that capture the number of data items (known as mass) within a subspace of streaming data. In this context, the mass is used to profile the degree of anomaly. The OzaBaggingADWIN and HS-Tree form an incremental hybrid IDS on the 6BR.

\subsubsection{Adaptivity}
\label{sec:Adaptivity}

Adaptive learning  updates the predictor model to respond to concept drift through the predictor operations. The 6LoWPAN  traffic routing evolves as nodes move or become unavailable (e.g. their energy resource may deplete), which results in reconstruction of the DODAG routing graph. Data forms a stream into the IDS  with a distribution that varies  over time.  To reduce memory use, concept-drift-based IDS trains over a small number of training data at any point in time and does not load the entire dataset into memory \cite{bhuyan2012survey}. 
The fundamental function of any concept drift detection approach is the mechanism to detect the drift occurrence timestamp. Accurate identification of the time that drifts happen plays a vital role in enhancing the system's adaptivity performance. Since the model never has full access to the entire data in a continuous environment, this article employs the adaptive sliding window (ADWIN) concept \cite{bifet2007learning} to perform concept drift detection. A window is a snapshot of data; it gives more importance to the recently observed data and periodically discards the older data. ADWIN observes the ratio of changes between two sub-windows ($W$) to compute the window size. When the difference between the simple mean ($\hat{\mu}$) of $W_{t}$ and $W_{t+1}$, differs more than the threshold $(\delta)$, the ADWIN algorithm concludes a drift is taking place and drops the oldest bucket and shrinks the window size; otherwise, no data is dropped and the window size increases by one \cite{bifet2007learning}.


\begin{figure*}[t!] 
\centerline{\includegraphics[width=15cm]{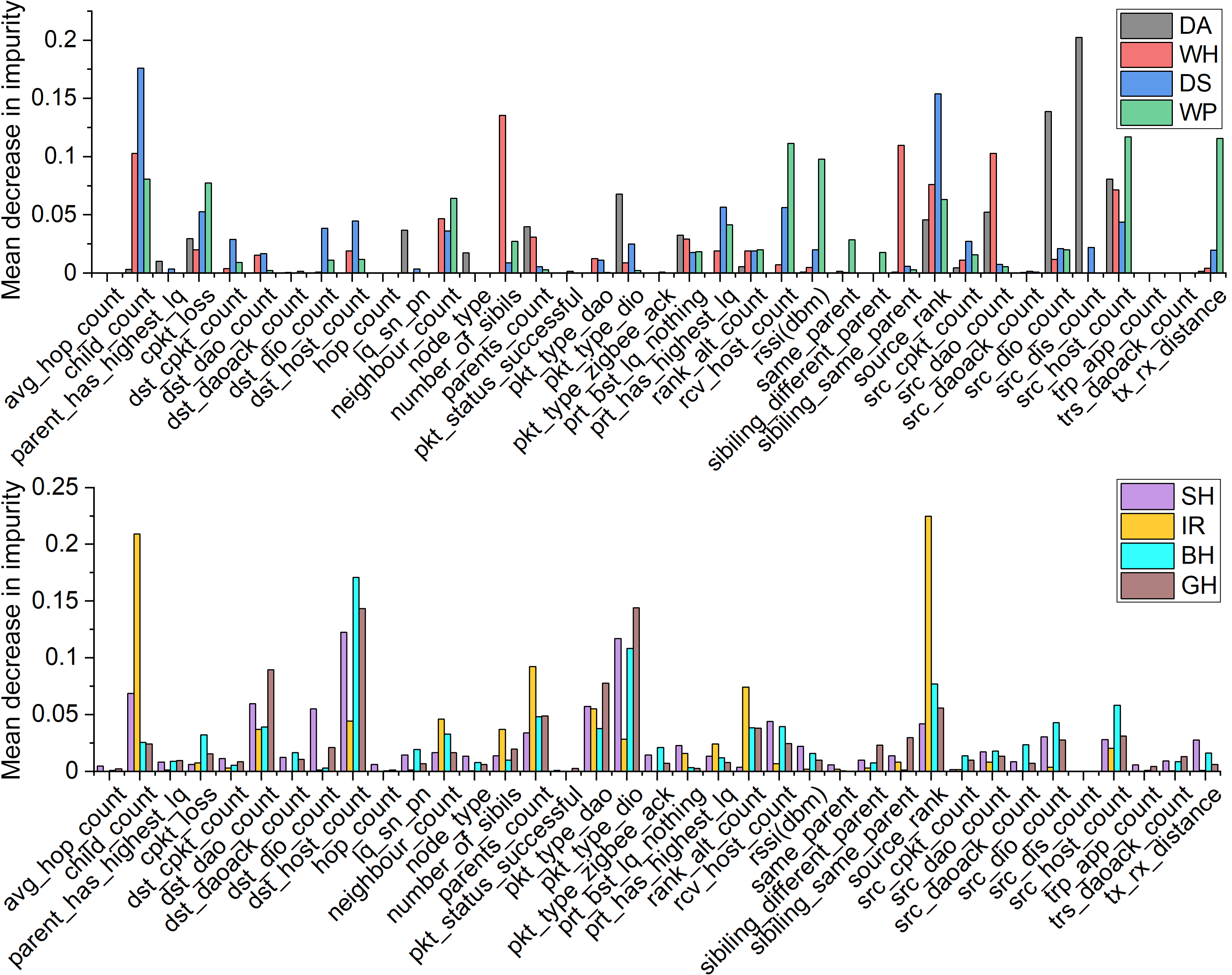}}
\caption{Feature Importance.}
\label{fig:feature_importance}
\end{figure*}

\begin{table}[htbp]
  \caption{Simulation Parameters.}
  \label{tab:simulation_parameters}
  
\begin{center}
\scalebox{1.}{
  \begin{tabular}{|l|l|}
\hline

    \textbf{Parameters} & \textbf{Values} \\ 
    \hline

    Simulator & Tetcos Netsim V13 \\ 
    \hline

    Number of nodes & 16, 32, 64, 128 \\ 
    \hline

    Number of Malicious nodes & $\sim$10\%, $\sim$20\%, $\sim$30\%  \\ 
\hline
    
    Number of Workstations & 4, 8 \\ 
\hline
    
    Transmission Range & 50m \\ 
\hline
    
    Number of ML detectors &  $\sim$10\% \\ 
\hline

    \emph{Number of Mobile nodes} & $\sim$20\% \\ 
    \hline
    
    Scenario Dimension (Terrain) & (250 × 250) to  \\ 

    & (850 × 850) s.meters\\
\hline

    Traffic Rate & 250 kbps \\ 
\hline
    
    Simulation time & $\sim$ 21,600 seconds \\ 
\hline
    
    Application Protocols & COAP, CBR \\ 
\hline
    
    RPL mode & Storing mode \\ 
\hline
    
    Mobility Modes & Random Walk, Group Walk \\ 
\hline
    
    Path Loss Model & Log Distance, Exponent(n): 2 \\ 
\hline
    
    Distance between nodes & 25 $\sim$ 45 m \\ 
\hline
    
    Objective Function (OF) & OF0, LQ \\ 
\hline
    
    Receiver Sensitivity & -85 dBm \\ 
\hline

\end{tabular}}
\end{center}

\end{table}

\section{Implementation and Evaluation}
\label{sec:Implementation and Evaluation}

In this article, we use the Netsim simulator to evaluate the performance of the proposed scheme against different RPL attacks. In this context we consider different network configurations (e.g. number of malicious and legitimate nodes, and objective function), as described in Table \ref{tab:simulation_parameters}. The simulated 6LoWPAN  scenarios include 16 to 128 LLN nodes (excluding 6BR and external computers), with 10\% to 30\% of nodes assigned as malicious. In all scenarios, we consider 20\% of the nodes, including half of the malicious nodes, are mobile and randomly move around the terrain with a velocity of 5 m/s. Nodes distribute over terrain covering $250m^{2}\sim800m^{2}$ and are $25\sim45m$  apart, \emph{with $50m$ transmission range}. Each scenario is simulated for $\sim$360 minutes for performance benchmarking. 
This article uses the interleaved test-then-train approach to evaluate the proposed scheme \cite{bifet2009new}. It is assumed that the packets in the streaming data $D$ sequentially appear in the target network, where ${x}_{t}$ is an unlabeled instance vector observed at time $t$, containing different attributes about the node configurations and the DODAG. The actual label $y_{t}$ of instance $x_{t}$ will be available to the system at different points in time. In the continuous data environment like 6LoWPAN, the ground truth $y_{t}$ may not be available immediately before observing $x_{t+1}$, and it may be available at some point in future \cite{bhuyan2012survey}. Additionally, the observations of the data stream in the 6LoWPAN are independent. That means there is no relation between $(x_{t+1}, y_{t+1})$ and $(x_{t}, y_{t})$.

  





    
    
    

    


    
    
    
    
    
    
    
    



\subsection{Data-set and Feature Construction}
\label{sec:Feature_Engineering}

The simulations generate a dataset $D$, representing  malicious and normal (safe) network communications. Each observation $x$ in $D$ denotes a set of $n$ features $x=\{f_{1}, f_{2} ... f_{n}\}$, where $f_{i}$ contains specific information about the sender and receiver. The header of each RPL control packet (e.g. DIO, DIS, DAO) contains different information about the sender of the packet \cite{rfc6550,rfc6551} that can facilitate the identification of anomalous network activities. Engineering a set of informative features is essential to develop an IDS to accurately classify all types of RPL attacks in the streaming data environment. Therefore, we perform feature engineering to facilitate the classification of data streams for IDS. The extracted features can enable the anomaly-based classifiers to correctly identify all the anomalies  through training over normal instances and make signature-based classifiers to accurately classify each type of RPL attack. The raw instances of 6LoWPAN simulations contain a set of features that are not applicable for conducting intrusion detection tasks. For instance, features that represent node identities (e.g. IP address, MAC address, and node id) can inhibit scheme generalisation. Since this article employs a passive decentralised monitoring approach \cite{mayzaud2017distributed}, any feature that requires the internal configuration of legitimate nodes (e.g. power consumption, geographical location, CPU/RAM/ROM usages) are excluded. We simulated several pairs of networks $(\mathcal{A}, \mathcal{B})$ where $\mathcal{A}$ contains only  the normal nodes and $\mathcal{B}$ contains both the normal and malicious nodes. Observing the statistical difference of control and application packets in $\mathcal{A}$ and $\mathcal{B}$ enable us to identify the adverse impact that each RPL attack has in the networks in $\mathcal{B}$.
A simulated 6LoWPAN includes legitimate (safe) network communications (control and application packets) and malicious traffic. In each RPL attack scenario, malicious nodes cause adverse impacts inside the network by either generating malicious network traffic (e.g. DIS flooding, DIO suppression, and sinkhole attacks) or modifying legitimate network communication of their neighbouring nodes (blackhole and grayhole attacks). The abnormalities that each RPL attack causes inside 6LoWPAN constitute malicious observations.

\begin{figure*}[t!]
\begin{subfigure}{.5\textwidth}
\centerline{\includegraphics[width=8cm]{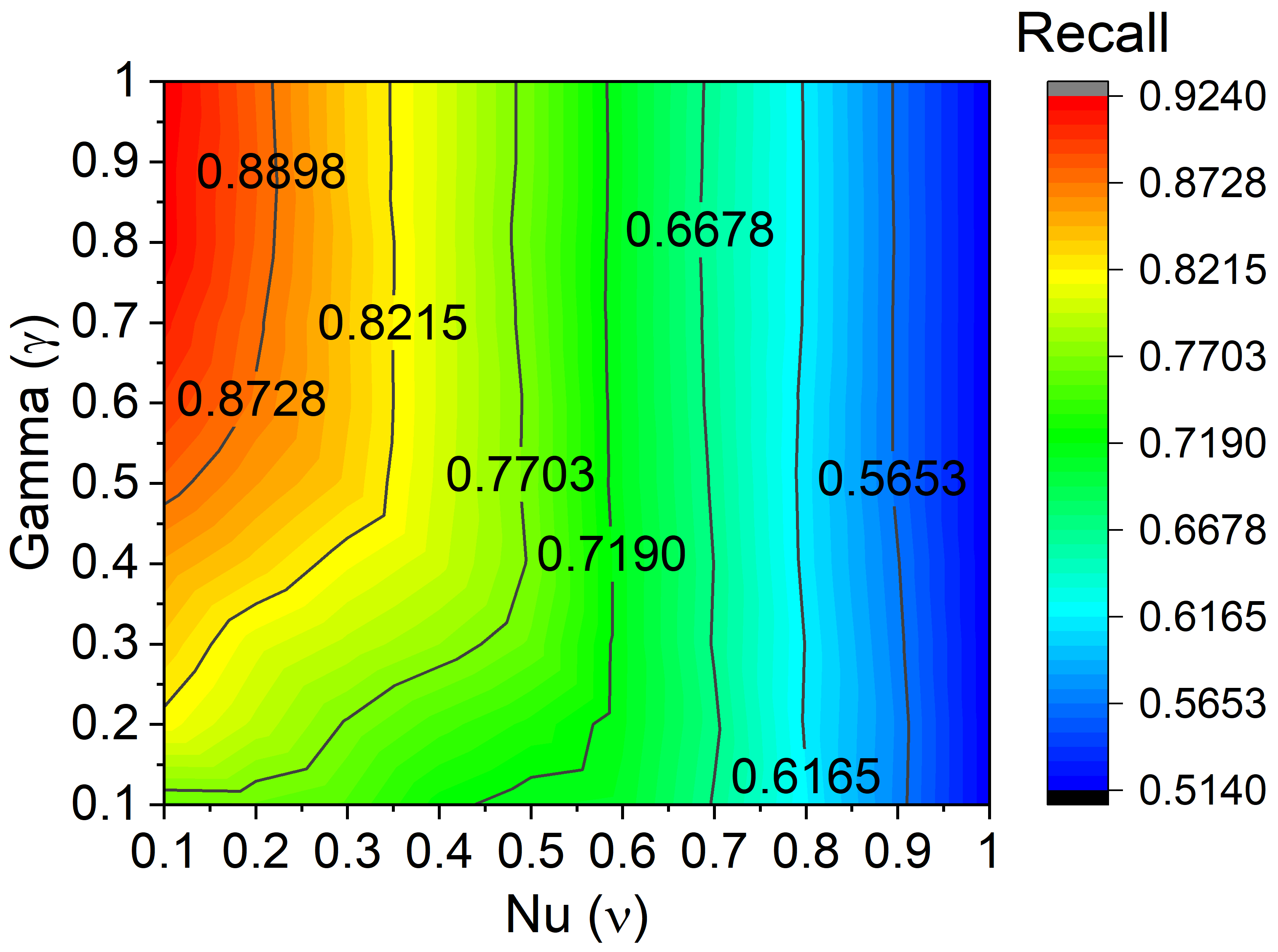}}
\caption{One-Class SVM Recall.}
\label{fig:OCSVM_Recall_3D}
\end{subfigure}%
\begin{subfigure}{.5\textwidth}
\centerline{\includegraphics[width=8cm]{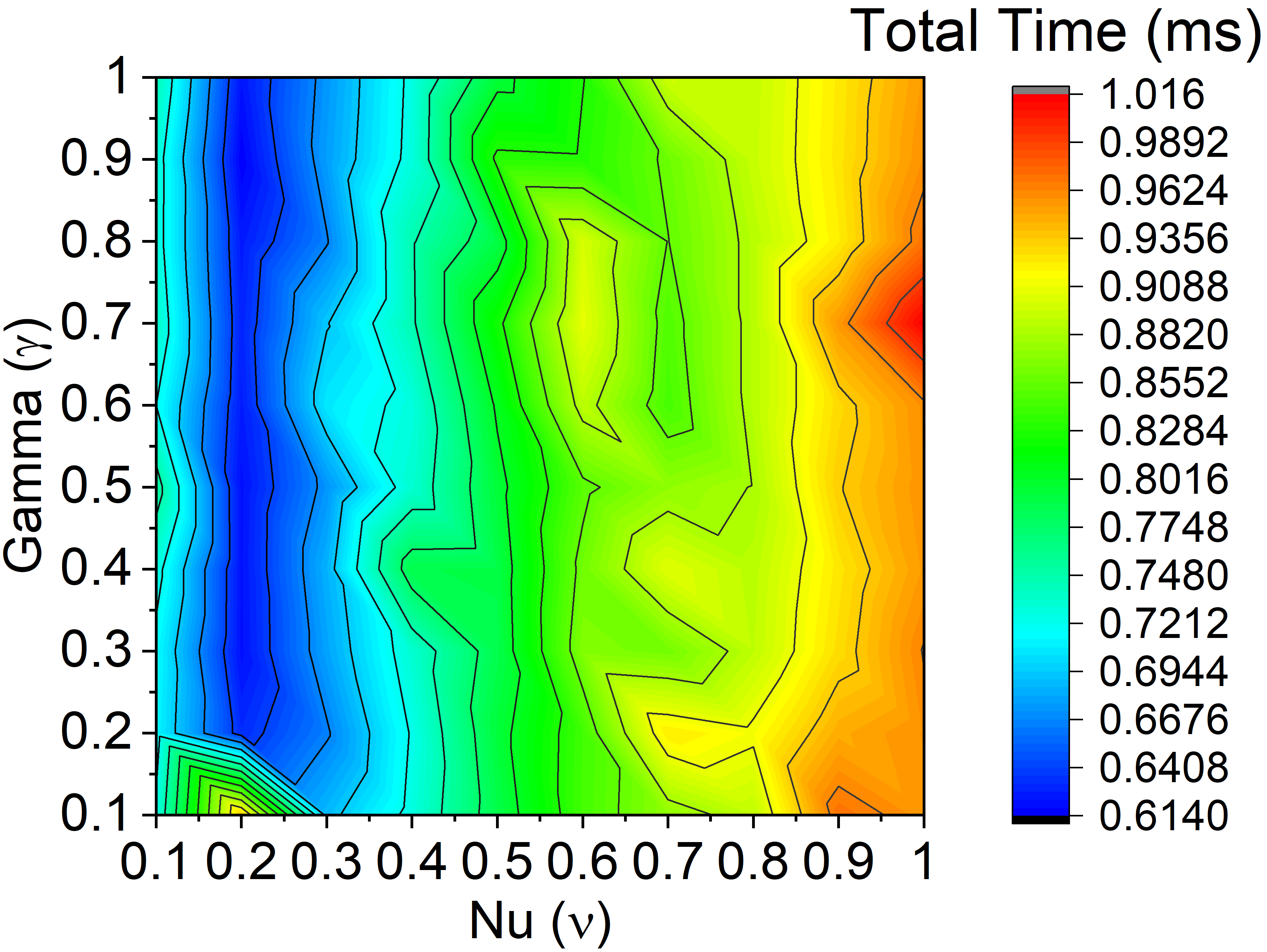}}
\caption{One-Class SVM Time Complexity.}
\label{fig:OCSVM_time}
\end{subfigure}
\caption{One-Class SVM (OCSVM).}
\label{fig:OCSVM_Result}
\end{figure*}

We extract three types of features: basic, time-based, and connection-based features. Basic features contain general node information derived from ICMP v6 control packet headers (node rank, source and destination addresses, flags etc). Whereas the time-based features provide information about the number of times that the current node sends or receives a specific type of application or control packet. Connection-based features carry salient information about the sender's routing configuration (RSSI, link quality etc) and the number of collided control and application packets perceived by an IDS detector. Table \ref{tab:Engineered_features} depicts the set of features engineered in this article. Here we apply the Mean Decrease in Impurity (MDI) importance metric to illustrate the importance of engineered features in identifying RPL attacks, as shown in Fig. \ref{fig:feature_importance}. The connection- and history-based features play vital roles in detecting the routing attacks in 6LoWPAN.


\begin{table}[t!]
\caption{Engineered Features.}

\begin{center}

\scalebox{.9}{
\begin{tabular}{|l|l|l|}
\hline

& \textbf{Feature} & \textbf{Description} \\ 
\hline

\multirow{4}{*}{\rotatebox[origin=c]{90}{\textbf{Basic}}} & pkt\_type & Type of packet (DIO, DAO, DIS, App etc) \\ 
\cline{2-3} 

& pkt\_status & Packet status (Collided, Successful) \\ 
\cline{2-3} 

& src\_rank & Sender rank in DODAG \\ 
\cline{2-3}


& adv\_vn & Advertised version number \\ 
\hline

\multirow{15}{*}{\rotatebox[origin=c]{90}{\textbf{History-based}}} & snd\_dis\_count & No. of DIS unicasted/multicasted by sender \\ 
\cline{2-3} 

& snd\_dio\_count & No. of DIO advertised by sender \\ 
\cline{2-3} 

& snd\_dao\_count & No. of DAO unicasted by sender \\ 
\cline{2-3} 

& snd\_daoack\_count & No. of DAO-Ack unicasted by sender \\ 
\cline{2-3}

& snd\_cpkt\_count & No. control packet issued by sender \\ 
\cline{2-3} 

& rcvd\_dis\_count & No. of DIS rcvd by current node in past \\ 
\cline{2-3} 

& rcvd\_dio\_count & No. of DIO rcvd by current node \\ 
\cline{2-3} 

& rcvd\_dao\_count & No. of DAO rcvd by current node \\ 
\cline{2-3} 

& rcvd\_daoack\_count & No. of DAO-Ack rcvd by receiver \\ 
\cline{2-3}

& rcvd\_cpkt\_count & No. of control packets rcvd by receiver \\ 
\cline{2-3} 
















& avg\_intpkt\_time & Average delay between pkts issued by snd\\ 
\cline{2-3} 

& rnk\_alt\_count & No. rank alteration by sender \\ 
\cline{2-3} 


& vn\_alt\_count & No. version number alteration by sender \\ 
\cline{2-3} 



& trans\_app\_count & No. of application trans by sender\\ 
\cline{2-3} 

& pkt\_e2e\_delay &  Packet end-to-end delay \\ 
\hline

\multirow{11}{*}{\rotatebox[origin=c]{90}{\textbf{Connection-based}}} & cpkt\_loss &  Control packet loss ratio  \\ 
\cline{2-3} 

& pkt\_loss &  Application packet loss ratio  \\ 
\cline{2-3} 

& avg\_hopcount & Average No. of hopcount (global view)  \\ 
\cline{2-3} 

& neighbour\_count & No. of neighbouring node \\ 
\cline{2-3} 

& child\_count & No. of children \\ 
\cline{2-3} 

& same\_parent & Sender and the detector have same parent \\
\cline{2-3} 

& rx\_sen  & Average receiver sensitivity  \\ 
\cline{2-3} 

& tx\_pwr & Average transmission power \\ 
\cline{2-3} 

& rssi & Received signal strength indicator of sender \\ 
\cline{2-3} 

& cmp\_snd\_prt\_lq & LQ of sender $>$ LQ of parent \\ 
\cline{2-3}


& prt\_bst\_lq & Current parent provide best link quality \\ 
\hline


\end{tabular}}
\label{tab:Engineered_features}

\end{center}


\end{table}

\subsection{Performance Evaluation and Discussion}
\label{sec:Performance_Evaluation}
As discussed in Section \ref{sec:ANIDS}, the novelty or anomaly  detectors of the proposed scheme work by observing the control packets of their neighbours; if the current observation is identified as anomalous,  it will be further analysed by the heterogeneous hybrid ensemble IDS on the 6BR. Below, different outlier detection, incremental ensembling, and concept drift detection algorithms are evaluated. We seek the best combination to gain the optimal F1, accuracy, recall, precision \cite{9383263} and kappa \cite{gomes2017survey,gama2014survey} with least False Negative Rate (FNR) and False Positive Rate (FPR) \cite{9383263}. Below, we conduct \emph{six} experiments utilising the underlying features of the Netsim emulator to execute the proposed framework over several Raspberry Pi 4 (model B, 4GB RAM) micro-controllers to measure the execution time and the model power consumption using a UM25C digital multimeter. Table \ref{tab:simulation_parameters} depicts the network configurations that we implemented to conduct our experiments. In all of our experiments, $\sim$ $20\%$ of nodes are mobile and randomly move around the terrain with a velocity of 5 m/s.

\begin{figure*}[ht]
\begin{subfigure}{.5\textwidth}
\centerline{\includegraphics[width=8cm]{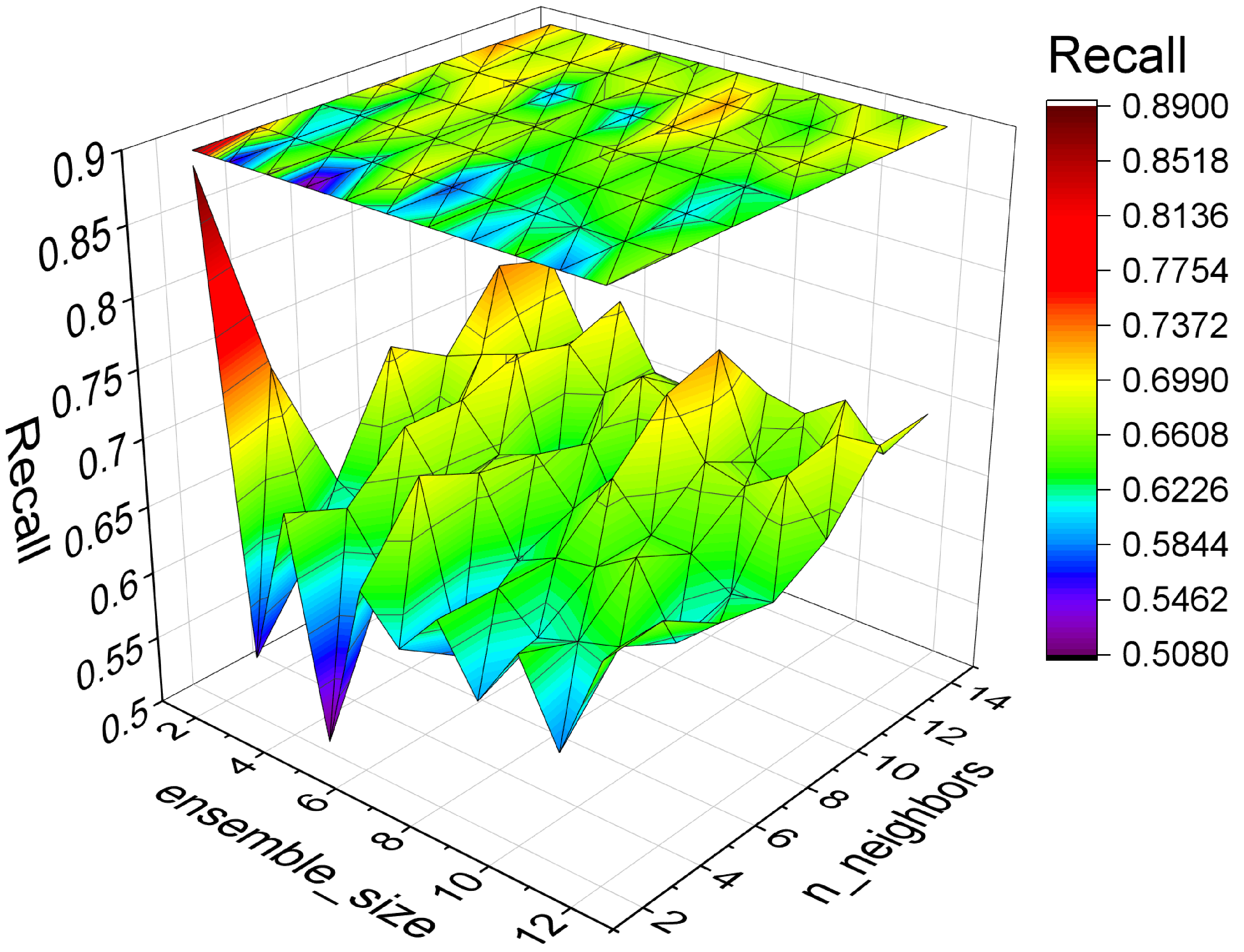}}
\caption{Local Outlier Factor Recall.}
\label{fig:LOF_Recall_3D}
\end{subfigure}%
\begin{subfigure}{.5\textwidth}
\centerline{\includegraphics[width=8cm]{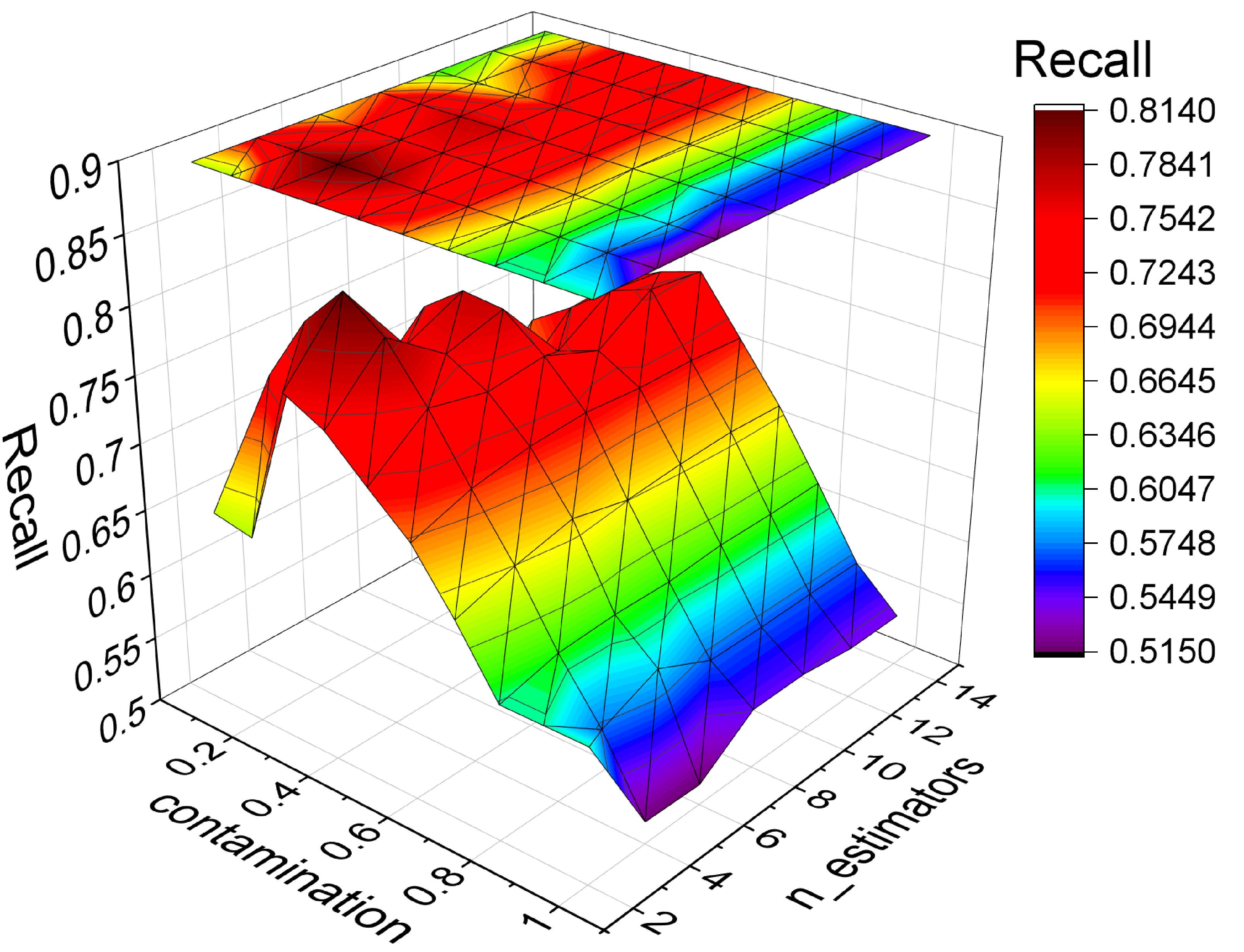}}
\caption{Isolation-Forest Recall.}
\label{fig:IF_Recall}
\end{subfigure}
\caption{Performance of different outlier detection algorithms.}
\label{fig:Outlier_algorithms_Recall_3D}
\end{figure*}

\textbf{Experiment 1.}\label{sec:Experiment One} The anomaly-based detector (also known as novelty detector) plays a crucial role to identify outliers in the proposed scheme. Here we measure the performance of OCSVM in detecting RPL attacks. We have evaluated OCSVMs with different parameter values for Nu $\nu$ $\in$ $(0,1]$ and Gamma $\gamma$ $\in$ $(0,1]$ for finding the optimal configuration; Fig. \ref{fig:OCSVM_Result}(a) shows that the OCSVM with $\nu \in (0.01, 0.25)$  and $\gamma \in (0.6, 1]$ can maximise recall. However, since the aim of the ANIDS is to identify all the intrusions and maximise TPR, here we assign the OCSVM with $\nu=0.2$ and $\gamma = 0.9$ to achieve 99.74\% TPR with 89.39\% recall (weighted average). Our experiments suggest that an OCSVM outperforms other existing anomaly detection algorithms, majority-voting ensemble of Local Outlier Factor and Isolation Forest, as shown in Fig. \ref{fig:Outlier_algorithms_Recall_3D}.



\begin{figure}[t!] 
\centerline{\includegraphics[width=9cm]{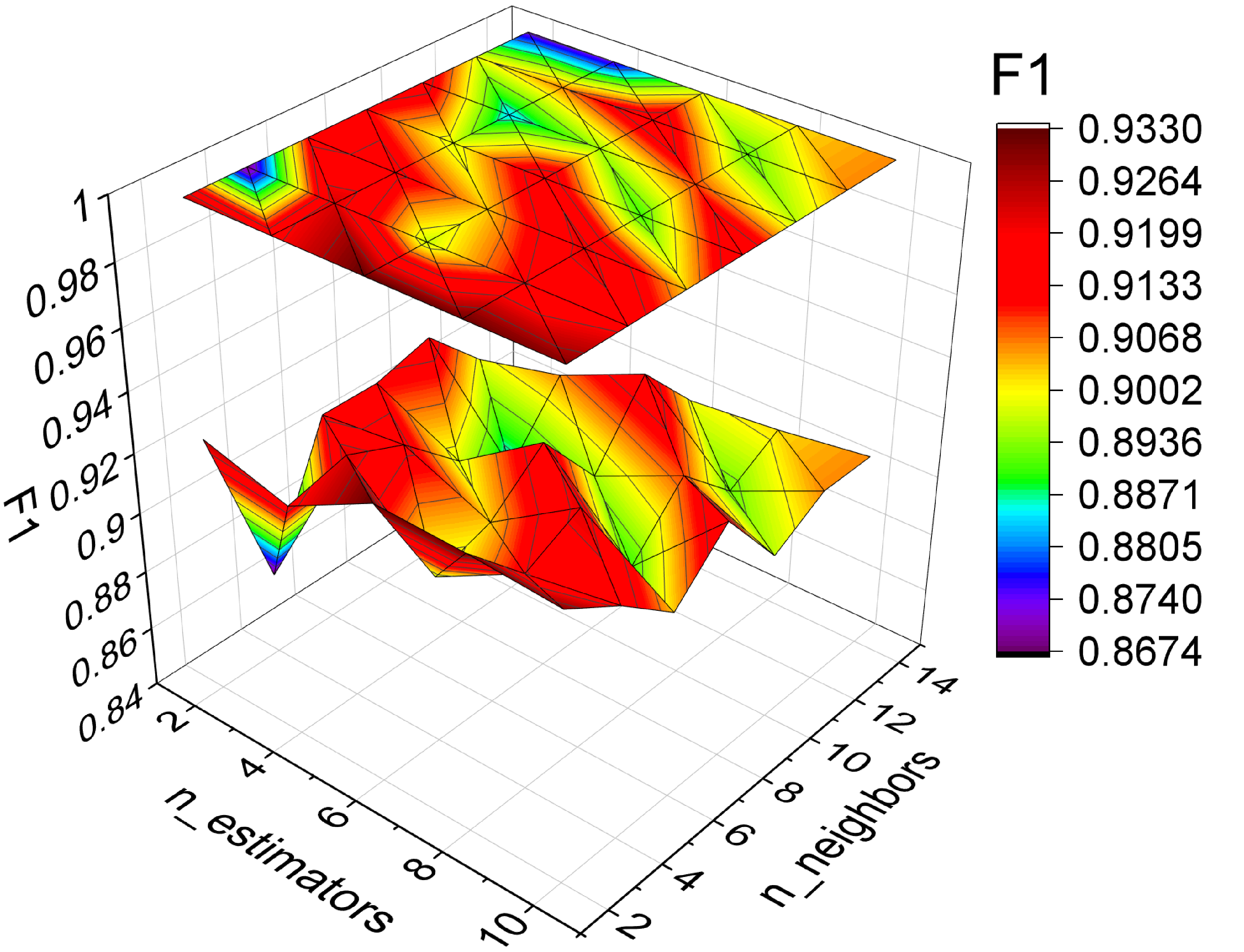}}
\caption{OzaBagging ADWIN (KNN) F1.}
\label{fig:OZABAGADWIN_KNN_3D_F1}
\end{figure}

\begin{figure*}[t!]
\centerline{\includegraphics[width=175mm]{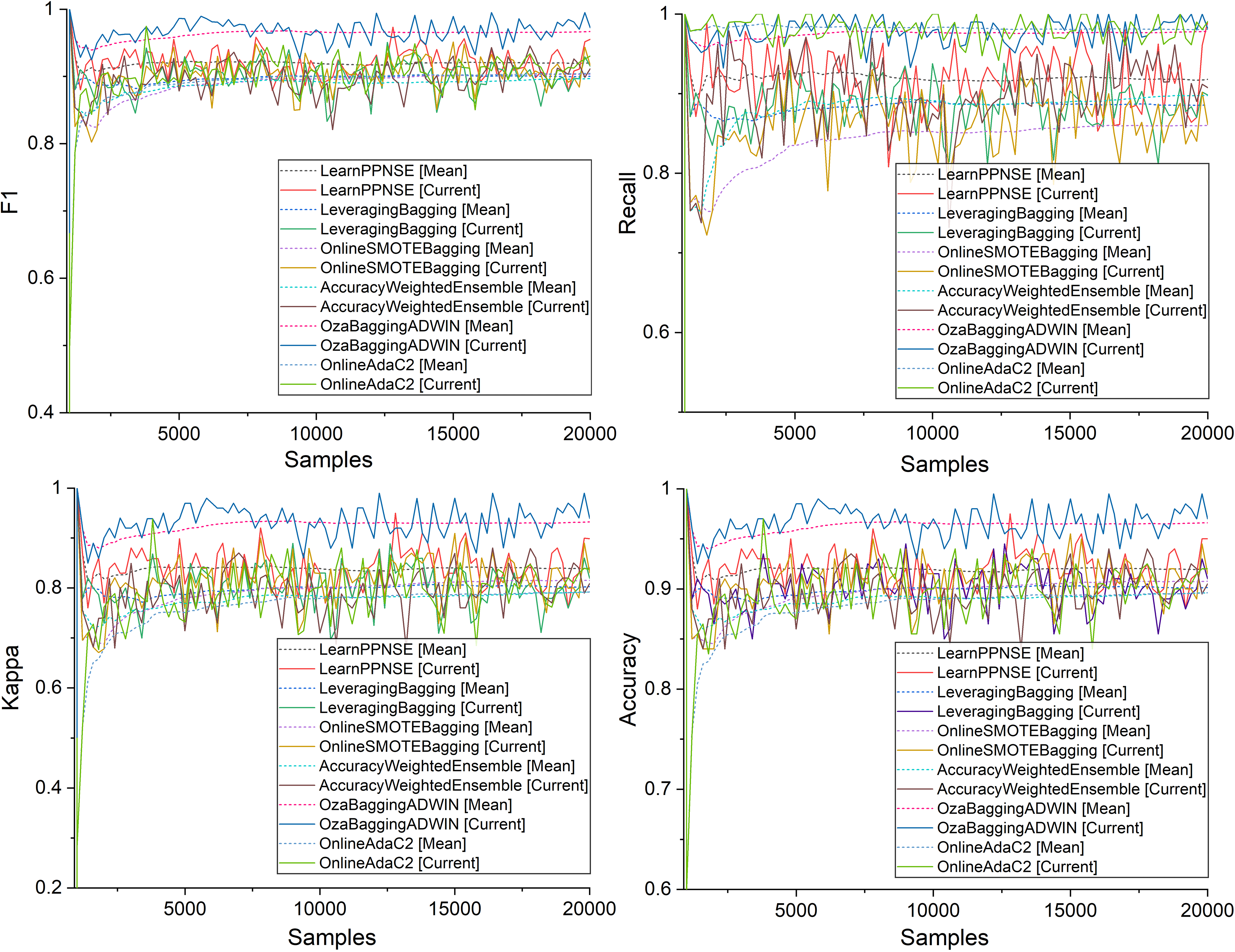}}
\caption{Performance of the proposed scheme in detecting RPL attacks, moving mean and current.}
\label{fig:Component_3}
\end{figure*}

\textbf{Experiment 2.}\label{sec:Experiment Two}  Experiment 1 showed that although the OCSVM algorithm can accurately identify outliers it also incurs 20.25\% FPR. To address this issue, we conduct our second experiment to  measure the performance of different incremental ensemble algorithms and rectify ANIDS mis-classifications. Here, we have compared the performance of OzaBagging \cite{oza2001online},  LearnPPNSE \cite{elwell2011incremental}, Online Boosting \cite{wang2016online}, Online AdaC2 \cite{wang2016online}, Accuracy Weighted Ensemble \cite{wang2003mining}, and Online SMOTE Bagging \cite{wang2016online} algorithms in detecting RPL attacks.
The outcome of our experiment (as shown in Fig. \ref{fig:OZABAGADWIN_KNN_3D_F1} and Fig. \ref{fig:Component_3}) shows that the combination of OzaBagging using KNNADWIN can provide the best possible outcome to identify known intrusions. OzaBagging using KNNADWIN with n\_estimators (number of estimators) as 4 and n\_neighbours (number of neighbours) as 6 receives 91.5\% F1 and 7.8\% FPR and with n\_estimators as 8 and n\_neighbours as 6 receives 92.2\% F1 and 7.3\% FPR, as depicted in Fig. \ref{fig:OZABAGADWIN_KNN_3D_F1}.



\textbf{Experiment 3.}\label{sec:Experiment Three} Above, we showed how  an incremental ensemble approach can identify known intrusions efficiently. Our proposed hybrid IDS targets both known and unknown intrusions. Accordingly, we now investigate an incremental ensemble of anomaly-based classifiers which can rectify false-negative alarms of the signature-based IDS. False-negative alarms are very costly and indicate the IDS failing in its primary task. In this experiment, we show how the inclusion of an incremental HalfSpace-Trees (HS-Trees) classifier can enhance the overall performance of the system. Fig. \ref{fig:Component_3} shows that the HS-Trees algorithm forms a better hybrid IDS  when it combines with the OzaBaggingADWIN compared to other incremental algorithms by around 6 to 10\%. Fig. \ref{fig:Component_3} gives the  current and moving mean (also referred to as moving average) F1, recall, kappa, and accuracy of the incremental ML algorithms.



\begin{figure*}[t!]
\begin{subfigure}{.5\textwidth}
\centerline{\includegraphics[width=8cm]{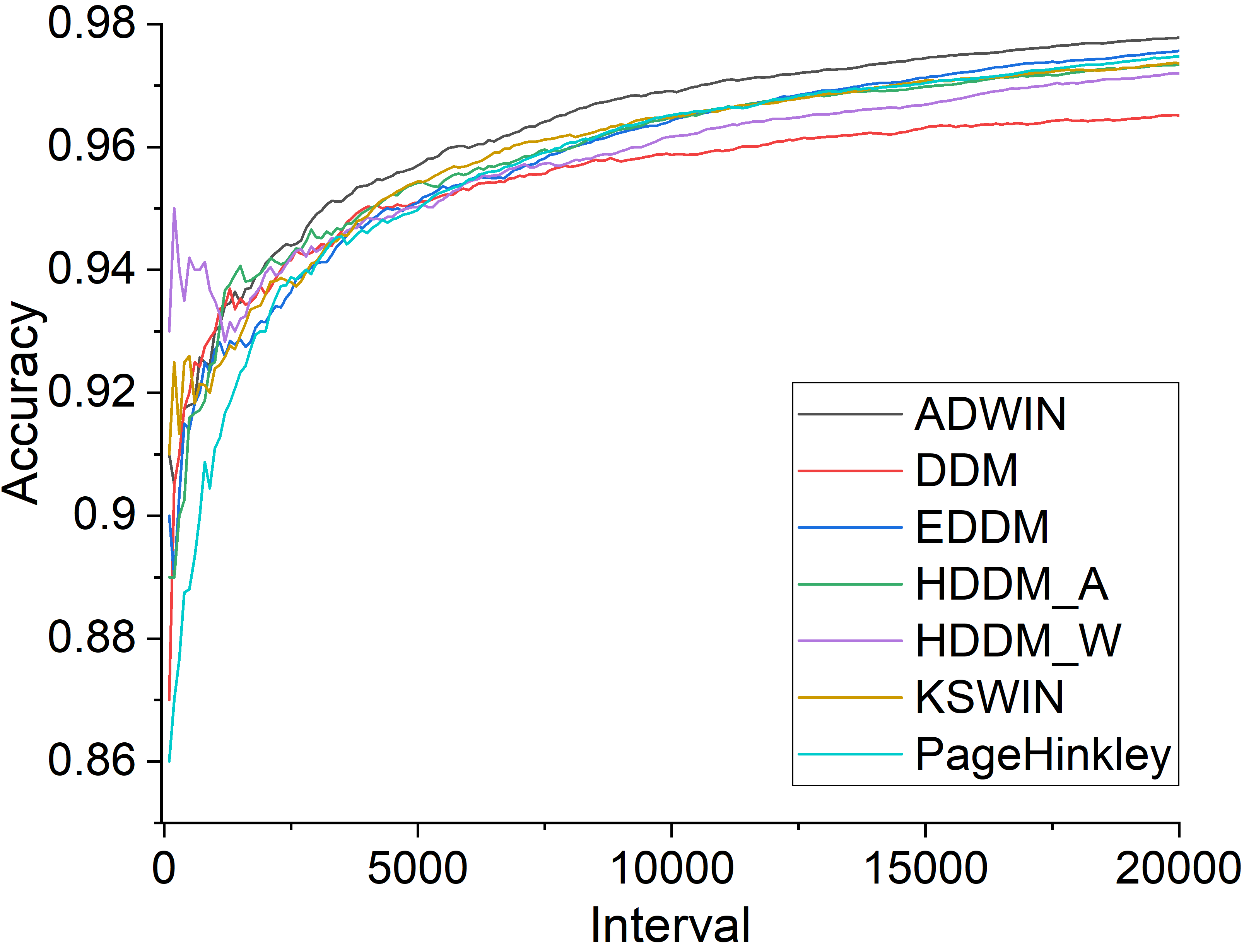}}
\caption{}
\label{fig:concept_drift_comparison}
\end{subfigure}%
\begin{subfigure}{.5\textwidth}
\centerline{\includegraphics[width=8cm]{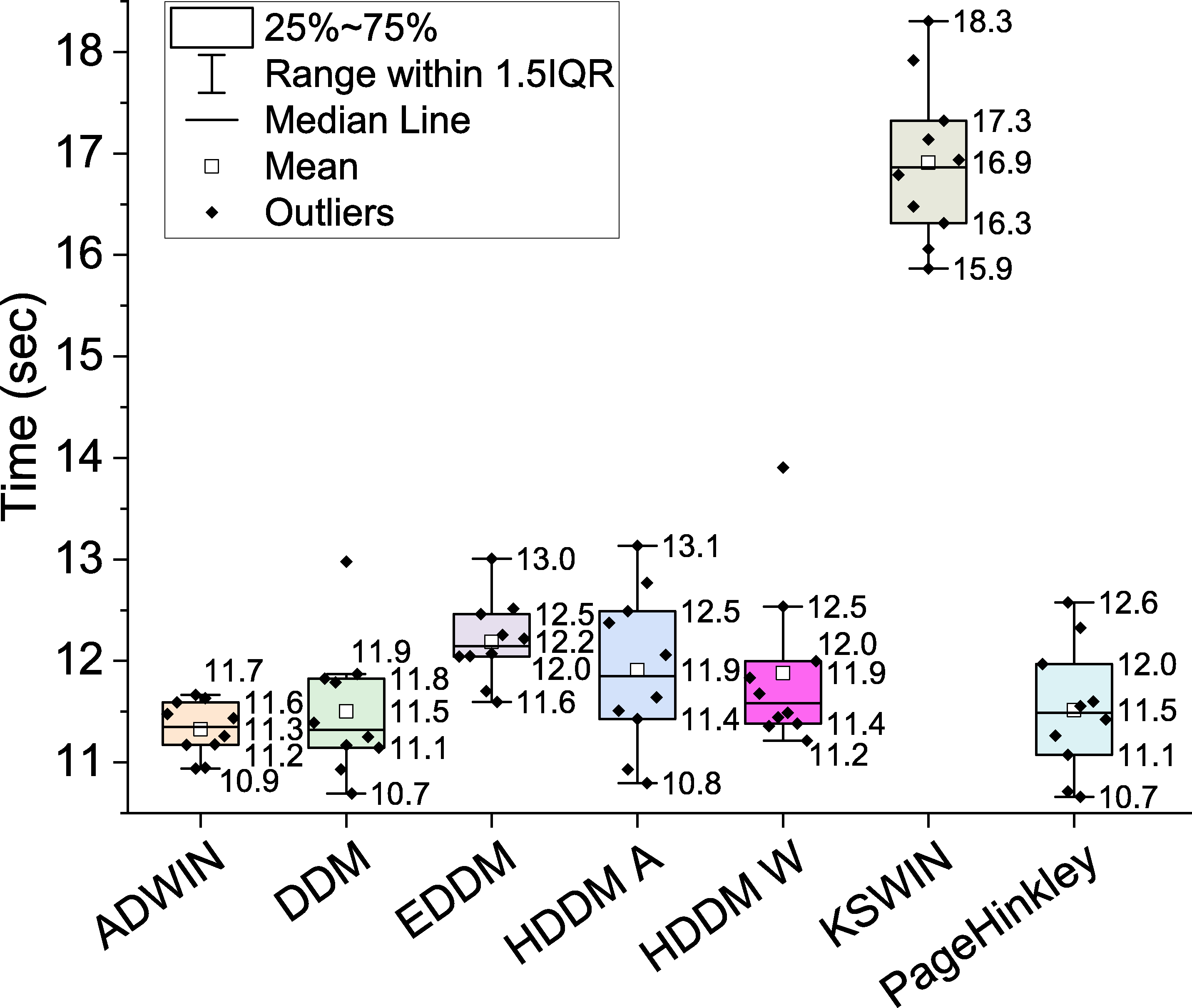}}
\caption{}
\label{fig:concept_drift_ACC_comparison}
\end{subfigure}
\caption{Comparison of Concept-drift Detection methods.}
\label{fig:Concept_Drift_Detection_Algorithms}
\end{figure*}

\textbf{Experiment 4.} \label{sec:Experiment Four}
Here, we investigate  to what extent concept-drift detection can provide system adaptivity. We evaluate different drift detection algorithms to select one that can ensure adaptivity in the system and also enhance the framework performance over time. We consider the following (seven) adaptive Windowing methods for concept drift detection: (ADWIN), Drift Detection Method (DDM), Early Drift Detection Method (EDDM),  Kolmogorov-Smirnov Windowing (KSWIN), PageHinkley, Drift Detection Method based on Hoeffding’s bounds (HDDM) with moving weighted average-test (HDDM-W) or moving average-test (HDDM-A) concept drift detection methods \cite{yuan2018concept,gama2014survey}. Results are presented in Fig. \ref{fig:Concept_Drift_Detection_Algorithms}(a) and Fig. \ref{fig:Concept_Drift_Detection_Algorithms}(b). From Fig. \ref{fig:Concept_Drift_Detection_Algorithms}, we can see that ADWIN gives the best accuracy than of the concept-drift detection methods in the shortest time interval. Outcomes of Experiments 1, 2, 3, and 4 show that our proposed scheme so far addresses \textit{DP1} (adaptive and robust intrusion detection which were discussed in Section \ref{subsec:Desirable_Properties}).


\begin{table*}[htbp]

\caption{Performance Bench-marking.}
\begin{center}
\scalebox{1.}{
\begin{tabular}{|c|c|c|c|c|c|c|c|c|c|c|c|c|c|c|c|c|c|}
\hline 
\textbf{N}&\textbf{M}&\multicolumn{8}{|c|}{\textbf{Accuracy}}&\multicolumn{8}{|c|}{\textbf{FNR}} \\
\cline{3-18} 
\textbf{} & \textbf{} & 
\textbf{\textit{SH}}& \textbf{\textit{BH}}& \textbf{\textit{GH}}& \textbf{\textit{DA}}& \textbf{\textit{IR}}& \textbf{\textit{WH}}& \textbf{\textit{DS}}&  \textbf{\textit{WP}} 

& \textbf{\textit{SH}}& \textbf{\textit{BH}}& \textbf{\textit{GH}}& \textbf{\textit{DA}}& \textbf{\textit{IR}}& \textbf{\textit{WH}}& \textbf{\textit{DS}}&  \textbf{\textit{WP}}
\\

\hline
\multirow{3}{*}{16} & 10\% & 91.5 & 91.8 & 96.2 & 99.8 & 95.8 & 98.3 &  97.4 & 98.6 & 14.1 & 13.8 & 3.4 & 0 & 7.3 & 2.4 & 2.6  & 2.7  \\ 
\cline{2-18}
 & 20\% & 98.7 & 95.4 & 98.4 & 100 & 97.9 & 96.5 & 98.7 & 97.5 & 1.8 & 5.4 & 3.0 & 0 & 4.0 & 4.4 & 2.3 & 2.9 \\ 
\cline{2-18}
 & 30\% & 97.6 & 97.0 & 96.6 & 100 & 94.1 & 99.6 & 98.2 & 99.5 & 3.4 & 5.4 & 5.8 & 0 & 11.3 & 0.1 & 2.9 & 0.2 \\ 
\hline 

\multirow{3}{*}{32} & 10\% & 93.3 & 96.3 & 98.5 & 99.8 & 97.8 & 99.7 &  98.5  & 99.6 & 10.0 & 5.4 & 2.0 & 0.3 & 3.8 & 0.2 & 2.2 & 0.5 \\ 
\cline{2-18}
 & 20\% & 98.7 & 98.2 & 98.2 & 100 & 97.8 & 94.8 & 98.4 & 95.2 & 2.4 & 3.1 & 2.0 & 0 & 3.6 & 9.5 & 2.0 & 8.7 \\ 
\cline{2-18}
 & 30\% & 98.6 & 98.3 & 98.7 & 100 & 97.0 & 90.1 &  98.7 & 91.9 & 2.3 & 3.2 & 2.4 & 0 & 5.3 & 16.0 & 2.4 & 13.2 \\ 
\hline 

\multirow{3}{*}{64} & 10\% & 92.5 & 93.1 & 90.6 & 99.9 & 94.9 & 91.6 & 89.5 & 92.7 & 13.8 & 13.2 & 16.8 & 0.1 & 9.1 & 12.0 & 18.9 & 10.6\\ 
\cline{2-18}
 & 20\% & 93.0 & 93.4 & 96.2 & 100 & 94.9 & 91.0 & 97.0 & 96.3 & 11.0 & 11.4 & 6.7 & 0 & 8.4 & 10.7 & 4.9 &  6.7\\ 
\cline{2-18}
 & 30\% & 93.7 & 93.8 & 96.2 & 100 & 96.4 & 94.5 & 98.7 & 96.6 & 11.5 & 9.4 & 7.1 & 0 & 5.0 & 10.1 & 2.4 & 5.9 \\ 
\hline 

\multirow{3}{*}{128} & 10\% & 97.2 & 93.0 & 91.2 & 99.8 & 95.5 & 93.5 & 94.0 & 92.3 & 5.4 & 13.4 & 16.0 & 0.4 & 8.1 & 9.2 & 8.2 & 11.3 \\ 
\cline{2-18}
 & 20\% & 93.6 & 93.9 & 94.1 & 100 & 95.9 & 94.4 & 96.0 & 93.1 & 11.7 & 11.0 & 10.0 & 0 & 6.1 & 10.5 & 6.7 & 13.3 \\ 
\cline{2-18}
 & 30\% & 94.3 & 94.9 & 96.9 & 100 & 96.9 & 95.2 & 96.7 & 95.4 & 10.0 & 8.4 & 5.8 & 0 & 4.7 & 8.5 & 5.8 & 7.8 \\ 
\hline

\multicolumn{18}{|c|}{\text{\textbf{SH:} Sinkhole; \textbf{BH:} Blackhole; \textbf{GH:} Grayhole; \textbf{DA:} DIS Flooding; \textbf{IR:} Increase Rank;}}\\
\multicolumn{18}{|c|}{\text{\textbf{WH:} Wormhole; \textbf{DS:} DIO Suppression; \textbf{WP:} Worst Parent; }}\\
\multicolumn{18}{|c|}{\text{\textbf{N:} Total number of nodes; \textbf{M:} No. Malicious nodes; No. Mobile nodes $\sim$20\%}}\\
\hline 

\end{tabular} 
}
\label{tab:IDS_performance_TPR_FNR}

\end{center}
\end{table*}


\begin{table*}[t!]
\caption{Performance Bench-marking with offline IDS in 6LoWPAN}
\begin{center}
\scalebox{.93}{
\begin{tabular}{|c|c|c|c|c|c|c|c|c|c|c|c|c|}
\hline
\textbf{Paper} &  \textbf{No.} & \textbf{No.} & \textbf{Duration} & \textbf{Mobility} &  \multicolumn{8}{|c|}{\textbf{RPL Attacks}} \\
\cline{6-13} 
 &  \textbf{Nodes} & \textbf{Malicious} & \textbf{minutes} & & \textbf{SH} & \textbf{BH} & \textbf{GH} & \textbf{IR} & \textbf{DA} & \textbf{DS} & \textbf{WH} & \textbf{WP} \\ 
\hline

\cite{bostani2017hybrid} &  5$\sim$50 & 1$\sim$5 & 20 & No & 100\% & - & 85.36\%  & - & - & - & 96\% & - \\ 
 &   &  &  &  &  &  & $\sim$92.68\% &  &  &  &  $\sim$97.53\% &  \\ 
\hline 

\cite{farzaneh2019anomaly} & 20$\sim$40 &  1$\sim$30\% & 30 & No & - & - & - & - & 100\% & - & - & - \\ 
\hline

\cite{foley2020employing} & 11 & 1 & 30 & No & 93.14\% & 93.14\% & -  & -  & -  & -  & -  & -  \\  
\hline

\cite{kasinathan2013ids} & 10 & 1 & - & No & - & - & - & - & - & - & -  & - \\ 
\hline

\cite{mayzaud2016using} & 2$\sim$10 & 1 & 480 & No & - & - & - & - & - & - & - & - \\ 
\hline

\cite{napiah2018compression} & 8 & 1$\sim$3 & $\sim$30 & No & 100\% & - & - & - & 100\%  & - & 100\%  & - \\ 
\hline

\cite{pongle2015real} & 8$\sim$24 & 1$\sim$2 & 30 & No & - & - & - & - & - & - & 94\% & - \\ 
\hline 

\cite{raza2013svelte} & 8$\sim$64 & 1$\sim$4 & $\sim$30 & No & 79\% & - & 81\% & - & - & - & - & - \\ 
\hline

\cite{shreenivas2017intrusion} & 4$\sim$8 & 2 & - & No & 90\% & - & - & - & - & - & - & - \\ 
 &   &  &  &  & $\sim$100\% &  &  &  &  &  &  &  \\ 
\hline

\cite{shukla2017ml} & 10$\sim$200 & $\sim$2 & - & No & - & - & - & - & - & - & 71\% & - \\
 &   &  &  &  &  &  &  &  &  &  &  $\sim$75\% &  \\ 
\hline

Proposed &  16$\sim$128 & 10$\sim$30\% & 360 & Yes & 91.5\% & 91.8\% & 90.6\% & 94.1\% & 99.8\% & 94.0\% & 90.1\% & 91.9\% \\ 
Scheme$^{\dagger}$ &   &  &  & (20\%) & $\sim$98.7\% & $\sim$98.3\%  & $\sim$98.7\% & $\sim$97.9\% & $\sim$100\% & $\sim$98.7\% & $\sim$99.7\% & $\sim$99.6\% \\ 
\hline


\multicolumn{13}{l}{\text{$^{\mathrm{*}}$results indicate the accuracy of the proposed IDS in detecting each type of RPL attack;}}\\

\multicolumn{13}{l}{\text{$^{\dagger}$ Details are shown in Table \ref{tab:IDS_performance_TPR_FNR};}}\\

\end{tabular}
}
\label{tab:benchmarking}
\end{center}

\end{table*}

\begin{table*}[t!]
\caption{Unknown Attack Detection.}
\begin{center}
\scalebox{1.}{
\begin{tabular}{|c|c|c|c|c|c|}
\hline
\textbf{Unknown}&\multicolumn{5}{|c|}{\textbf{Performance Metrics}} \\
\cline{2-6} 
\textbf{Attack} & \textbf{\textit{Accuracy}}& \textbf{\textit{Precision}}&  \textbf{\textit{F1}} & \textbf{\textit{TPR}} & \textbf{\textit{FPR}}  \\
\hline
SH & 90.85 & 91.16 & 90.79 &  86.52  & 5.17  \\
\hline
BH & 89.75 & 90.30 & 89.74 & 83.62 & 3.55  \\
\hline
GH & 93.9 & 94.07 & 93.88 & 90.97 & 3.31 \\
\hline
IR & 91.75 & 92.20 & 91.71 & 86.61 & 3.25 \\
\hline
DA & 98.30 & 98.36 & 98.29 & 96.57 & 0  \\
\hline
WH & 98.35 & 98.36 & 98.34 & 97.04 & 0.30  \\
\hline
DS & 93.95  & 94.05 & 93.94 & 91.62 & 3.76  \\
\hline
WP & 95.10 & 95.18 & 95.09 & 92.93 & 2.71   \\
\hline
\end{tabular}
}
\end{center}
\label{tab:unknown_detection}%

\end{table*}


\textbf{Experiment 5.}\label{sec:Experiment Five} Here, we measure the time complexity of each component in the proposed framework. We consider 64 LLN nodes in 6LoWPAN, with 20\% assumed malicious. Fig. \ref{fig:OCSVM_Result}(b) shows the results over 1500 network packets, where 50\% of instances are assumed normal and the remaining 50\% include each RPL attack type equally. 
We measure the time complexity for each ANIDS and CIDS separately. Fig. \ref{fig:OCSVM_Result}(b) shows the time complexity that the OCSVM with  $\nu =0.2$ and $\gamma =0.8$ causes the least time complexity in the system. On the other hand, the adaptive heterogeneous hybrid IDS, developed in our Experiments 2 and 3, using 4 learners and 8 neighbours (KNN) causes $O(log(n))$ time complexity in the system. Table \ref{tab:Time_Complexity} shows that ANIDS has linear and logarithmic time complexity in training and testing, while CIDS has polynomial time complexity in the proposed scheme. To measure the power consumption of each component, we use the Netsim Emulator feature to connect the physical micro-controllers with the simulation environment and connect digital ammeters to the micro-controllers. We run our experiment for 10 minutes, disabling all unnecessary background tasks and applications. The power consumption of an ANIDS and the CIDS in a LLN with 64 nodes was 3.505 J/s and 3.754 J/s, respectively, whilst a legitimate node without any ANIDS or CIDS consumed 3.17 J/s. In this way, we have satisfied \textit{DP2} (lightweight IDS).


\begin{table}[htbp]
\caption{Time Complexity.}\label{tab:Time_Complexity}
\begin{center}

\scalebox{1.}{
\begin{tabular}{|l|l|l|}
\hline
\textbf{Comp} & \textbf{Training (sec)} & \textbf{Testing (sec)}\\
\hline
ANIDS  &  $O(N)$: & $O(log(n))$: \\
 & 0.36 + -2.4E-08*n & 0.22 + -0.0021*log(n)\\
 \hline
CIDS & - & $O((log$ $n)^{k})$: \\
&  & -2.3 * $x^{0.94}$\\
\hline
\end{tabular}}
\end{center}

\end{table}

\textbf{Experiment 6.}\label{sec:Experiment Six}
Here, we first evaluate how well the proposed scheme detects each RPL attack in LLNs with different proportions of legitimate and adversarial nodes, while 20\% of nodes, including 50\% the malicious nodes, were mobile and moving, as shown in Table \ref{tab:IDS_performance_TPR_FNR}.

From Table \ref{tab:IDS_performance_TPR_FNR}, we can see that the performance of the proposed scheme is plausible in terms of the accuracy and false-negative rate (FNR) for detecting various RPL attacks. The proposed scheme can detect IR attack with high accuracy (up to $\sim$97.9\%); and the SH, BH, DS, and GH attacks with up to $\sim$98.7\%; WH with up to $\sim$99.7\%; WP with up to $\sim$99.6\%; and DA with up to $\sim$100\% accuracy. Our outcomes show that our proposed scheme satisfies \textit{DP3} (accurate in evolving data environment) and \textit{DP4} (detect a wide range of RPL attacks). 
We then consider the detection of unforeseen intrusions, where each RPL attack was excluded from the pre-training data one-by-one and exclusively covered all the adversarial activities of the evaluation data stream, as shown in Table \ref{tab:unknown_detection}. Outcomes of this experiment show our proposed scheme can address \textit{DP5} (detect unseen/unknown intrusions).

\section{Conclusion}
\label{sec:Conclusion}
Routing threats in 6LoWPAN and threats against RPL are highly significant. In this article, we have introduced an adaptive hybrid heterogeneous IDS scheme that is effective and efficient and can readily cope with changes to the environment and detect known and unknown routing intrusions in the 6LoWPAN. Table \ref{tab:benchmarking} gives an \emph{indicative} comparison between our scheme and the results obtained by other authors. However, we stress our results are obtained in a much more challenging environment. We provide our results here as a bench-mark for the research community.


\bibliographystyle{IEEEtran} 
\bibliography{mybibliography}

\end{document}